\documentclass[11pt,a4paper]{article}
\usepackage[utf8]{inputenc}
\usepackage[inner=2.5cm,outer=2.5cm, top=2.5cm, bottom=2.5cm]{geometry}
\usepackage{authblk}
\usepackage[english]{babel}

\usepackage{mathptmx}
\usepackage{amsmath}
\usepackage{array,multirow,makecell}
\usepackage{amsfonts}
\usepackage{amssymb}
\usepackage{graphicx}
\usepackage{caption}
\usepackage{transparent}
\usepackage{cancel}
\usepackage{url}
\usepackage{float}
\usepackage{xfrac}
\usepackage[dvipsnames]{xcolor}
\usepackage[caption = false]{subfig}
\usepackage{textgreek}
\usepackage{setspace}
\usepackage{pgfplots}
\usepackage{varwidth}
\usepackage{import}
\usepackage{lipsum}
\pgfplotsset{compat=1.14}
\usepackage{mathrsfs}
\usepackage{lmodern}
\usepackage{textcomp}
\usepackage{siunitx}
\usepackage{multirow}
\usepackage{indentfirst}
\usepackage[justification=centering]{caption}
\usepackage[pdftex]{hyperref}
\hypersetup{colorlinks=true,linkcolor=black,citecolor=Blue,urlcolor=black}
\usepackage[capitalise,noabbrev,nameinlink]{cleveref}
\title{Recommendation for a Standard Rolling Noise Machine}
\author[1]{M. Edwards\thanks{matthew.edwards.@matelys.com}}
\author[2]{R. Gonzalez Diaz}
\author[3]{N. Dallaji}
\author[1]{L. Jaouen}

\affil[1]{Matelys Research Lab, 7 Rue des Maraîchers, Bât B, 69120 Vaulx-en-Velin, France}
\affil[2]{Aalto University, Department of Computer Science, P.O. Box 13000, 00076 Aalto, Finland}
\affil[3]{Moelven Töreboda AB, Bruksgatan 8, 545 31 Töreboda, Sweden}

\date{July 2020}

\begin{document}

\maketitle

\begin{abstract}
In the world of building acoustics, a standard tapping machine has long existed for the purpose of replicating and regulating impact noise. However there still exist other kinds of structure-borne noise which could benefit from being considered when designing a building. One of these types of sources is rolling noise. This report details a proposal for defining a standard rolling noise machine. Just as the standard tapping machine can be used in any building and on any surface as a way of characterizing and comparing the performance of various floors with respect to impact noise, the development of a standard rolling device would enable the same evaluation and comparison to be made with respect to rolling noise. The hope is that such a prototype may serve as a launch pad for further development, spurring future discussion and criticism on the topic by others who may wish to aid in the pursuit of a truly standardized rolling noise machine.
\end{abstract}

\section{Introduction} \label{sec:RollingMachine_introduction}
In the world of building acoustics, a standard tapping machine has long existed for the purpose of replicating and regulating impact noise. This device was originally designed to mimic the sound of human footfall \cite{reiherUberSchallschutzDurch1932}, and while this is indeed the strongest source of annoyance among survey responders (in regards to strictly floors) \cite{Jarnero962744}, there still exist other kinds of structure-borne noise (whose primary transfer path is through the floor) which could benefit from being considered when designing a building. One of these types of sources is rolling noise.

\begin{figure}[!htbp]
	\centering
	\includegraphics[width=\textwidth]{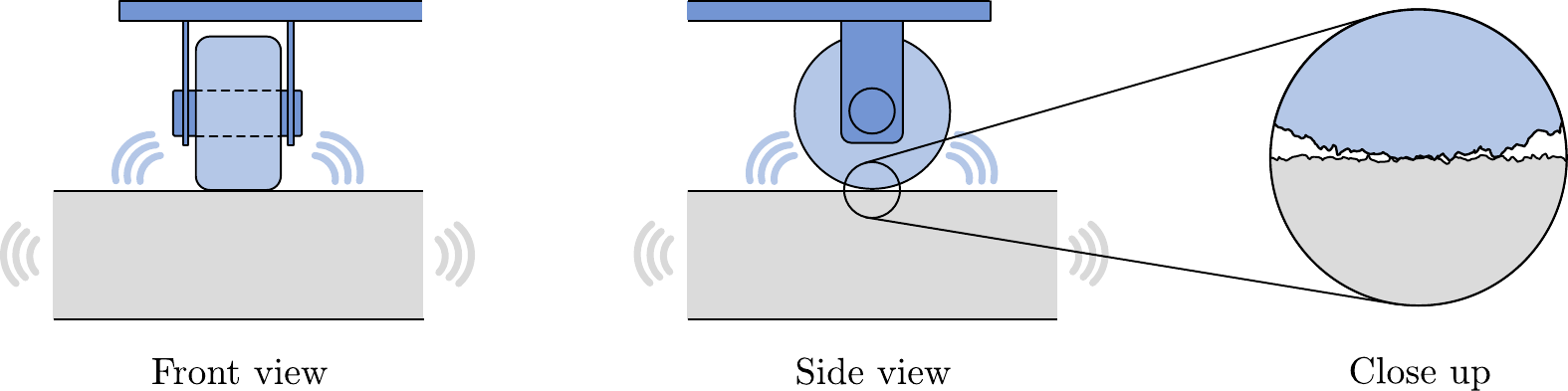}
	\caption{A typical rolling noise schematic. The small-scale roughness between the floor and wheel generates structural vibrations as the wheel rolls across the floor.}
	\label{fig:contact_schem}
\end{figure}

When it comes to assessing the performance of flooring construction in multi-story buildings, the tapping machine is by far and away the most widely used device. This use is justified, to be sure, as impact noise is often given as one of the greatest sources of annoyance to multi-story building inhabitants \cite{forssenAcousticsWoodenBuildings2008}. Other sources of noise exist, however, such as rolling noise. This is often talked about in regards to outdoor sources such as trains and automobiles, but there are a plethora of indoor rolling sources which can cause annoyance for building inhabitants as well: delivery trolleys in commercial spaces, rolling desk chairs in offices (both personal and commercial), children’s toys, and suitcases, for example. As shown in \cref{fig:contact_schem}, these items generate noise due to the small-scale roughness between the floor and wheels, which causes structure-borne noise to propagate through the building as they roll across the floor.

\begin{figure}[!htbp]
	\centering
	\includegraphics[width=0.6\columnwidth]{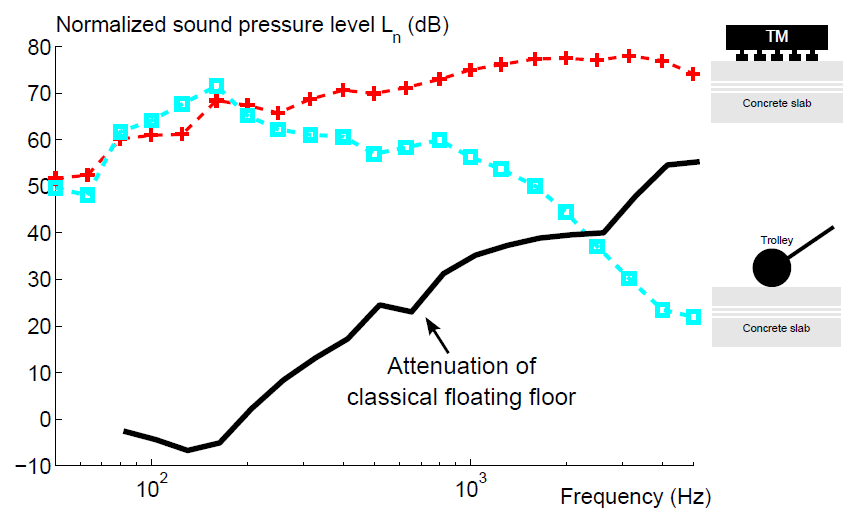}
	\caption{Comparison of the spectra of tapping noise and rolling noise of a classical concrete floor, as well as the attenuation of a classical floating floor (140~mm concrete slab + a decoupling layer + 40~mm screed). \cite{edwardsRollingNoiseModeling2018}}
	\label{fig:spectra_comp_app}
\end{figure}

\cref{fig:spectra_comp_app} shows the spectra of a typical tapping machine and rolling noise on a classical concrete floor, as well as the attenuation of a classical floating floor. The sound signature of impact noise is quite different than that of rolling noise in both the temporal and spectral domains. Considering the prevalence of tapping machines, this mismatch means that focus is rarely given to the effects of rolling noise when designing acoustic treatment systems for floors, resulting in a gap in performance. There is no guarantee that the techniques which are developed to reduce impact noise will necessarily be effective at reducing rolling noise. Furthermore, without a repeatable way of replicating and measuring this kind of noise, the processes to go about finding solutions to it will remain impeded and incongruent.

This report details a proposal for defining a standard rolling noise machine. Just as the standard tapping machine can be used in any building and on any surface as a way of characterizing and comparing the performance of various floors with respect to impact noise, the development of a standard rolling device would enable the same evaluation and comparison to be made with respect to rolling noise. It should be noted that, should the proposal be carried forth to the development of a prototype itself, this device should be treated as just that: a prototype. The development of a rolling noise machine that is robust enough to satisfy all the necessary requirements of being deemed ``the standard'' is no simple task. As such, one should not expect to accomplish such a lofty goal on the first attempt. Rather, the hope is that such a prototype may serve as a launch pad for further development, spurring future discussion and criticism on the topic by others who may wish to aid in the pursuit of a truly standardized rolling noise machine. If such a device is intended to be as to rolling noise as a tapping machine is to impact noise, then analyzing how the tapping machine came to be should serve as a good way to inspire its development.

\section{The standard tapping machine: A history} \label{sec:tapping_machine_history}
The purpose of this section is not to present a comprehensive overview of the history of the tapping machine. In fact, something of the sort has already been accomplished by Theodore J. Schultz in his extensive report ``Impact Noise Testing and Rating – 1980'' \cite{schultzImpactNoiseTesting1981}. On the contrary, this section shall discuss the aspects of the development of the tapping machine which have been deemed relevant to providing insight into how one may go about developing an equivalent device for rolling noise. The history of how the tapping machine become the standard for impact noise may be used as inspiration for how to go about setting a standard for rolling noise.

\subsection{The tapping machine as it exists today} \label{sec:tapping_machine_today}
\begin{figure}[!htbp]
	\centering
	\includegraphics[width=0.5\columnwidth]{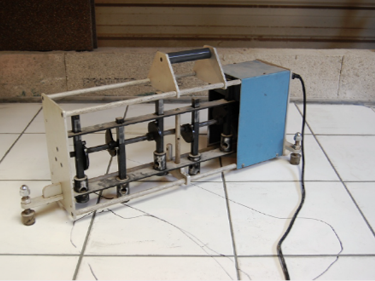}
	\caption{A typical tapping machine. The five hammers repeatedly drop in succession to generate impacts with the floor at a rate of 10~Hz.}
	\label{fig:tapping_machine_app}
\end{figure}

The modern tapping machine, a schematic of which is shown in \cref{fig:tapping_machine_app}, is defined by ISO 10140-5 \cite{ISO10140520102010}. It has remained essentially unchanged since its first recommendation in 1960 \cite{FieldLaboratoryMeasurements1960}. It consists of five hammers, controlled by an AC motor on a cam system, which rise and fall in succession to generate impact events with the floor upon which the machine rests. The hammers each have a mass of 500 g (± 2.5\%), a fall height of 4~cm (± 2.5\%), and are dropped once every 100 ±5 ms. Each hammer is supposed to strike the floor only once each time it is dropped (though this has been shown to not always be the case in reality \cite{schultzImpactNoiseTesting1981}. The hammer material may be either brass or steel, with a third option to use a rubber hammer head (the properties of which are tightly controlled) on fragile floor surfaces. The hammer heads are to be 3~cm in diameter with slightly rounded faces, with a radius of curvature of 50~cm.

\subsection{How we got to now} \label{sec:how_we_got_to_now}
The story of how the tapping machine as we know it today came to become the standard for measuring impact noise in buildings is a long and winding one. Starting in 1927 with the first recorded study of footfall noise at the National Bureau of Standards in the United States, and reaching international agreement in January 1960 with the first adoption of the standard by the International Organization for Standardization (ISO), one may hasten to call the process by which the design of the tapping machine was settled upon to be 10\% scientific and 90\% political. Schultz summed up the major issue that was at hand rather nicely when he stated:
\begin{quotation}
``It is worthwhile to recall here that, in acoustical testing, we sometimes have to choose between making worthless measurements or making meaningless measurements. For example, in the case of impact noise: If we do our measurement in the field, using some kind of real-life excitation such as walking, the resulting impact sound level may be so low compared to the background noise that it cannot be accurately measured; therefore, the measurement is worthless. On the other hand, if we hammer harder (as with the hammer machine), so that the impacts can be readily measured, then the measurement is meaningless, because it tells us nothing about how the floor behaves under actual use.'' \cite{schultzImpactNoiseTesting1981}
\end{quotation}
Many of the decisions that were made in the design of the tapping machine erred towards being worthwhile (what exactly this means will be discussed later). As a result, the final product is often regarded by critics as yielding results which are meaningless. Nonetheless, the tapping machine we have is the tapping machine that we have: it is the device for measuring impact noise in buildings, and while there exists much contention over its usefulness, there still appears to be some agreement that it is still better to have a poor standard than no standard at all.

\subsubsection{Physical machine construction} \label{sec:machine_construction}
The mid 1900’s saw a plethora of different tapping machines being developed and tested by various teams across Europe and North America. Most of these drew inspiration from one another in some form, and as the years progressed, a regression towards the mean developed, as the various tapping machines became more and more similar, finally culminating in the standard we know today. Even from the very first impact test, the design has changed remarkably little. The original tapping machine, developed by the United States National Bureau of Standards, consisted of five rods which could be controlled by a DC motor with a cam system to rise and fall at separate times, roughly once every fifth of a second \cite{chrislerTransmissionSoundWall1929}. From there, other teams developed their own versions of the device; with the next significant event coming in 1938, when Germany standardized their own tapping device (DIN 4110, 1938) \cite{DIN411019381938}. The differences across these devices were found in the varying hammer masses, drop heights, hammer head materials, and frequency of impacts. However, the general construction and methodology (i.e. a device which automatically raises and lowers hammers for impacting the floor) remained essentially unchanged from the start. Perhaps this is not terribly surprising, as it is a relatively straightforward and intuitive way to generate impacts remotely while conducting measurements in the room below. Nevertheless, it begs the question of whether other designs were not given credence simply because of the strong mentality of ``this is the way we have always done it'' that tends to overpower group decision making, especially in a bureaucratic scenario such as that of writing an international standard.

\begin{figure}[!htbp]
	\centering
	\includegraphics{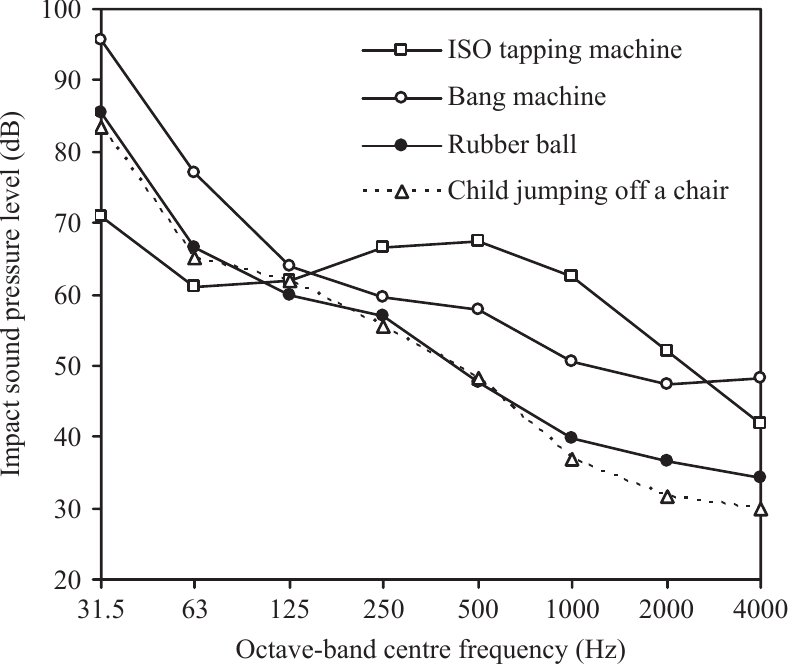}
	\caption{Impact sound pressure level on a timber floor from various impact devices, including the standard tapping machine and rubber ball. Reproduced from \cite{hopkinsSoundInsulation2007}.}
	\label{fig:tapping_vs_ball_spectra}
\end{figure}

As a matter of fact, other variations in the tapping machine design were indeed developed and tested. Furthermore, methodologies for reproducing impact noise which have nothing to do with the tapping machine at all have even been developed, some of which are currently in use today in some parts of the world. Perhaps the most well-known example of this is the rubber ball test (Also defined in ISO 10140-5 \cite{ISO10140520102010}), which is used in some eastern countries like Japan in place (or in addition to) the tapping machine, due to its ability to better replicate the sound of bare feet walking across a floor (a more common occurrence in some eastern cultures). Just as the name implies, this involves dropping a standardized rubber ball of mass 2.5 kg and diameter 18~cm from a height of 1 m multiple times, and recording the resulting impact sound in the room below. It can also be used in scenarios where low frequency impact performance is of concern, as it generates an impact noise which has more acoustic energy in the low frequencies than the tapping machine, as exemplified in \cref{fig:tapping_vs_ball_spectra} \cite{hopkinsSoundInsulation2007}.

In setting out to investigate the story of how the tapping machine came to be, we began with some sort of an expectation that the development process would have been rooted in deep mathematical analysis. We were hoping to find some sort of documentation throughout the research process which detailed the complex theoretical methodologies employed by these early researchers to develop the various tapping machine iterations that were designed throughout history: with vast numerical explanations which depicted how they arrived at their final prototypes. The reality was surprisingly (or perhaps not at all, depending on one’s familiarity with the matter) nothing of the sort. As far as we could find, the early tapping machines were developed almost completely by trial and error, with researchers simply trying different materials, weights, drop heights, and methodologies, until they found the one that sounded the best when listened to from the room below. There was a small amount of theoretical analysis involved, to be fair. For example, two parameters, fall energy (e.g. in \cite{NationalPhyscialLaboratory1935,gastellSchalldamnMessungenFraxisUnd1936,schuleWarmeundSchalltechnischeUntersuchungen1937}) and hammer momentum (e.g. in \cite{hofbauerSchallschutzDeckenSound1934,bauschSenalldammungsmessunnenImLaboratorium1939}), were discussed in several of these early designs. Fall energy is usually expressed in terms of gram-centimeters, and is defined simply by multiplying the hammer mass by the hammer fall height. Hammer momentum on the other hand is defined by multiplying the hammer mass by the square root of the fall height. These values were discussed, and a bit of argument went back and forth over the years over which one was more accurate at describing the loudness of a given impact noise (with hammer momentum eventually winning out and being accepted as the more correct term \cite{bauschSenalldammungsmessunnenImLaboratorium1939})). However, by far and away, these machines appear to have been designed experimentally. It is also worth noting that some of the early test devices were sparse in their details, neglecting to specify aspects of the test procedure (e.g. \cite{meyerSchallisolationUndSchallabsorption1930}), or even making no mention of whether or not they were designed to replicate human footfall noise (e.g. \cite{meyerSchallschutzHochbautenSound1931}).

The variations to the tapping machine itself that were tested during this gestation period mostly came in the form of differing hammer materials, impact frequencies, and fall energies. The most common materials used started out as wood, rubber, and leather, then migrated towards brass and steel as the years progressed. Anywhere from between one and six hammers were used, impacting at frequencies ranging from 4~Hz to 10~Hz. Hammer masses tended to be in the range of 100--1~000 g \cite{schultzImpactNoiseTesting1981}. Hammer drop heights observed the greatest range of variability, mostly due to the experimentation that was done with measurement procedures which involved hammer machines with variable drop heights (more on that in the following section).

\subsubsection{Measurement procedures} \label{sec:measurement_procedures}
The divide between low and high fall energy tapping machines is linked to the types of measurement procedures which were tried throughout its infancy. The procedure used today involves placing the tapping machine on the floor of the room above and measuring the sound pressure level in the room below. Thus, the performance depends solely on the sound level in the reception room. However, many early researchers experimented with using a transmission loss style approach, where the sound in both the emission and reception rooms was measured (e.g. in \cite{chrislerTransmissionSoundWall1929,MeasurementSoundNoise1939,lindahlMeasurementImpactSound1940}), and the difference between the two used to categorize the performance of the floor. Eventually researchers did move towards only measuring the sound level (or loudness level) in the reception room. Comparative tests do exist today: typically for evaluating the performance of a floor covering. Here the sound pressure level in the reception room is measured with and without the floor covering installed on a concrete slab, and the difference between the two calculated as the sound attenuation. Nevertheless, the measurements typically remain conducted solely in the reception room.

Subjective listening was also used in place of objective sound level or loudness measurements in many early studies. This was partially due to the fact that accurate objective measurement equipment had not really been invented yet, but also due to the belief that subjective measurements would yield results closer to reality. The two most common subjective methods were the reference sound method and the just audible method. In the reference sound method, the listener listened to the tapping machine sound in the reception room and compared it to a calibrated reference sound. This calibrated reference sound could have been anything from a pure tone (the frequency content of which was often never stated in these early reports) to an impulse caused by passing a voltage spike through a loudspeaker, with the voltage itself measured as the quantifying value. Examples of this can be found in \cite{NationalPhyscialLaboratory1935} and \cite{meyerSchallschutzHochbautenSound1931}. The just audible method involved using a tapping machine capable of dropping the hammers from varying heights, and having the listener listen to the impact events over several of these height cycles. By counting the number of impacts heard between periods of silences (i.e. when the impacts occurring were below the threshold of audibility), they could identify the lowest audible drop height. Examples of this can be found in \cite{reiherUberSchallschutzDurch1932,hofbauerSchallschutzDeckenSound1934,osswaldMethodMeasuringSound1936,kipferInsulationMeasurementFootstep1949}. It is worth noting that in essentially all of these early subjective tests, there was still a degree of calibration involved in the listening. Listeners did not judge the impact sound while standing in the reception room directly, but listened to it instead through an earpiece which was transmitting the sound from the reception room, and which had been calibrated so that the loudness of the sound through the receiver was perceived to be the same as the loudness of the impacts while standing directly in the reception room \cite{schultzImpactNoiseTesting1981}.

In the wake of World War II in the late 1940’s, a group of research teams met from England, France, Denmark, and the Netherlands to compare test procedures for impact noise \cite{kostenComparativeImpactSound1949}. They used the same kind of tapping machine and the same test procedure in each of their respective labs, but with different measurement equipment (i.e. they each used the measurement equipment available at their respective labs, and did not try to expressly ensure they were identical in calibration). The results were surprising, deviating 10--15 dB from one another, even after correcting for the absorptions of the various lab reception rooms. A second round of comparative testing was then done, this time using a single tapping machine in a single lab, but each of the research groups bringing their own measurement equipment to perform the same measurement procedure with. This time results were much closer to one another, staying within a range of ±2.5 dB after absorption correction. Nevertheless, such a range in the very same lab with the very same tapping machine was still found to be alarming to the researchers themselves, and a cause for concern about the validity of the tapping machine as a reliable test apparatus \cite{schultzImpactNoiseTesting1981}.

\subsubsection{Final approval} \label{sec:final_approval}
Despite the concerns raised, representatives from Denmark, France, the Netherlands, Sweden, and the United Kingdom met in 1948 to discuss adopting an international standard tapping machine. They agreed to use the already existing German standard tapping machine (defined in DIN 4110 1938 \cite{DIN411019381938}), with the only major change being to use brass rather than wood hammers, with an alternative option to use rubber hammers on fragile floors. This decision was made based on a desire to reduce as much as possible the variability in measurements, as the material properties of brass and rubber can be controlled tighter than wood \cite{schultzImpactNoiseTesting1981}. The recommendation was brought to a symposium of 32 countries later that year, with comments on the symposium code being submitted in 1949. A final document on the recommendation for an international tapping machine was approved shortly thereafter at a meeting in Copenhagen \cite{parkinProvisionalCodeField1949}. Schultz also points out the peculiarity that the German standard was agreed upon being recommended internationally despite the fact that no representative from Germany was present at the 1948 meeting. This could perhaps lead credence to the ``this is the way we’ve always done it'' theory being a powerful driver in the decision-making process that ultimately resulted in an international standard being adopted in 1960 which was based on a design created in 1938, despite the fact that numerous alternatives were developed, tested, and proposed during the two decades that separated the two.

As bureaucratic dealings tend to go, the process was still rather slow in reaching the ISO. The recommendation first came before a committee of the ISO in 1955, with the final draft being voted upon in 1958. Of the twenty member countries, this final draft was approved by sixteen, receiving three abstentions and one opposition (from Canada). Finally, in January 1960, the ISO officially adopted ISO standard R 140 for conducting impact noise testing (among other things), cementing the tapping machine’s standardization in international history \cite{schultzImpactNoiseTesting1981}.

\subsection{Teach the controversy?} \label{sec:teach_the_controversy}
Today there are acousticians who strongly criticize the use of the tapping machine as an effective means of replicating impact noise in buildings. This is not a recent phenomenon, as criticisms of the tapping machine have existed since the very beginning \cite{marinerCommentsImpactNoise1963,marinerCriticismISOImpact1964}. Alternatives have been proposed \cite{schultzAlternativeTestMethod1976a,loverdeDualratingMethodEvaluating2017,schollImpactSoundInsulation2001}, but nothing has been adopted. Schultz points out however, that at the time the German DIN 4110 1938 standard was made, we already knew essentially all of the same problems that we know today, yet they still decided to make the tapping machine the way that they did. This does not mean that criticisms of the tapping machine are moot; perhaps it instead serves as evidence to how heavily politics influence the decision-making process. In regards to using live walkers as a way to more accurately create standard footfall noise, Schultz states: ``It is assumed without question in each study that any machine at all would be preferable to using live people to excite the test floor (Only in Sippell’s paper (1932) \cite{sippellSchallschutzDurchBaukonstruktionsteile1940} is there a suggestion that live walkers were used, and even then the evidence is far from certain.).'' Ultimately, the international symposium in 1948 decided to do what they did because they wanted a way to be able to compare lab tests to one another across locations, and the standard they came up with, however lacking, was the best way to do it.

\subsection{Design criteria of a standard rolling noise machine} \label{sec:design_criteria}
In this section we shall draw on what was learned from the development of the tapping machine to inspire the conditions for a standard rolling noise machine. At its most basic level, a standard rolling noise machine should perform a simple function: it should roll across the floor of the emission room while generating a noise that can be measured in the reception room below. There are countless ways by which this can be achieved. The goal here is not to come up with the design for the prototype itself, but rather to propose a methodology for creating one, and present a discussion of what the priorities should be in the engineering process in order to ensure that, when the times comes, the device that ends up being developed is worthwhile. 

\subsection{Lessons learned from the tapping machine} \label{sec:lessons_learned}
So, with all of that being said regarding the history of the tapping machine, how does any of it help inform the decision on how to go about designing a standard rolling noise machine? Looking at the history of the development of the tapping machine, some trends start to emerge. First and foremost, there is a clear preference for repeatability over accuracy. If the device yields wildly different results when tested repeatedly in the same scenario, then its usefulness as a standard is hardly justifiable. Thus, a rolling device should follow suit. A rolling device rolling across a standard concrete floor in one lab should yield similar measured results to the same device rolling across a similar concrete floor in a different lab.

A second preference can be seen for a machine which requires little or no modification from location to location. Today’s tapping machine generates impacts which, for a lightweight floor, have the capacity to be quite a bit louder than real footfall noise. If one wishes to have a device which is standard across all floor environments, this is an unavoidable reality. However, this has been deemed to be preferable to the alternative: a device which replicates real footfall noise perfectly, but is therefore so quiet when being tested on high impedance floor structures that the results provide no beneficial meaning. The same should be true of a standard rolling machine. The device should be capable of generating rolling noise levels which are sufficiently above the background noise level for a wide range of flooring constructions, from thick concrete to lightweight timber.

During the development of the tapping machine, much focus was placed on choosing the right hammer material. This is a consideration that depends not only on sound, as a leather tipped hammer will generate a slightly different impact noise than a brass tipped one, but also on repeatability (linking it with the first point made above). Brass, steel, and rubber can be controlled much easier than wood or leather, making them more appropriate (in the eyes of the 1948 committee) as materials to be used in an international standard device. When looking at rolling noise, such a consideration is equally as important, if not more. The wheels of such a device should be chosen such that they can be used repeatedly, and for a long lifetime, without degrading to the point of significantly changing the noise being produced. This may prove difficult, as the rolling noise depends on the roughness of the wheel, so any change in the wheel surface will result in a change in sound. Though perhaps the use of a sufficiently hard material, such as brass or steel, may be able to mitigate these problems to an acceptable degree. This also has the added benefit of generating a higher overall sound level, reducing the likelihood that the results will be too close to the background noise level.

Perhaps the most important lesson that can be learned from the development of the tapping machine is that, no matter what the case, sacrifices will always have to be made. In an ideal world, one could create an impact noise device which perfectly replicates footfall noise in every possible scenario, while also being repeatable, easily measurable, and standardized. We do not live in an ideal world: concessions will always be necessary. However, a device which is perhaps less accurate than one would like is better than no device at all. After the end of the second world war, acousticians in Europe were eager to establish standards for noise in buildings as quickly as possible, as they did not want buildings which were being constructed to be done so hastily, and with no regards to acoustic quality \cite{kostenComparativeImpactSound1949,cremerStandForschungIm1949,dalpaluUseSemanticDifferential2017}. Furthermore, they recognized that ``more reliable test methods would have to be developed to permit widespread testing of new buildings'' \cite{schultzImpactNoiseTesting1981}. Today, the time-sensitive pressure to develop standards quickly no longer exists, but the need for test methods which can be used easily in widespread cases remains. A device which is so complex or so modular that it is rendered no longer useful for quick validation tests is hardly worth developing, even if it may satisfy the desires of the most purist of acousticians. With rolling noise, just as with impact noise, the difficulty lies in choosing which corners are worth cutting for the sake of simplicity and which ones are worth preserving for the sake of accuracy. 

\subsection{Specific considerations for the standard rolling noise machine} \label{sec:specific_considerations}
The parameters which dictate the impact sound of a tapping machine are the hammer mass, hammer material, hammer drop height, frequency of impacts, and floor construction. With rolling noise, the governing parameters are the roughness of the floor and wheel, material properties of the floor and wheel, floor construction, speed of the device, and load on the device. In both cases, the parameters related to the floor are the dependent variables of the equation, as the whole purpose is to assess their acoustical performance. This may cause an issue for rolling noise. Because the rolling noise depends not only on the intrinsic material properties and the macro construction of the floor, but also on its surface roughness, two identical floors with different surface finishes can yield different noise levels for the same source rolling across them. For example, in the case of rolling office chairs, subjective testing has shown there to be a noticeable difference in the sound produced when changing the flooring material upon which the chair rolls \cite{dalpaluUseSemanticDifferential2017}.

\begin{figure}[!htbp]
	\centering
	\includegraphics{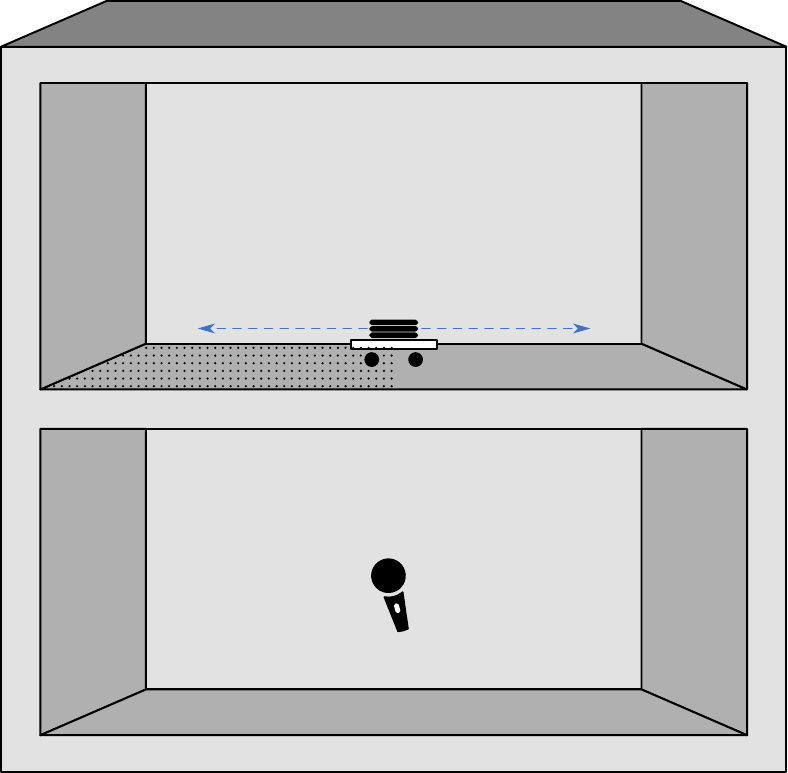}
	\caption{Example of a simple two-story concrete structure with varying surface roughness on the top floor.}
	\label{fig:two_story_varying_roughness}
\end{figure}

As an example, suppose we have a basic two-story structure with a simple concrete slab separating the two floors, such as the one shown in \cref{fig:two_story_varying_roughness}. In the west half of the room, the concrete remains as it was when it was poured (i.e. relatively rough). In the east half of the room, a polisher was used to polish the surface of the concrete during drying, resulting in a much smoother surface. The two halves of the floor remain identical in material composition and in construction, all that has changed is the surface roughness. In this scenario, a trolley rolling across the floor from east to west will exhibit a change in sound profile as soon as it crosses from rolling on the smooth concrete to rolling on the rough concrete (or vice versa). A tapping machine, on the other hand, will sound the same on both surfaces. All else being equal, whether this difference in rolling noise could be large enough to be detectable by a listener in the room below, or to raise concern for the feasibility of a standard rolling device, is yet to be known. Nevertheless, it is something that will need to be taken into consideration in the design process.

In turning next to the wheel (or rather, wheels) which will be used on this rolling device, the obvious question is that of what material they should be. While it is true that no single wheel material will be able to encompass the entire realm of possibility, the goal here is to replicate the ``worst case scenario'', developing a device which generates as loud a rolling noise as possible (so that levels on massive floors are still sufficiently high). To this end, a metallic wheel would likely be the best candidate. Alternatively, just as the tapping machine standard has the option for a rubber hammer to be used on fragile surfaces, perhaps two types of wheels may be specified for a rolling device: one soft and one hard.

It has been observed through rolling noise measurements that the presence of wheel flats can dominate over all other influencing factors in rolling noise. A wheel with flat spots rolling on a soft floor covering will generally still be louder than a smooth wheel rolling on bare concrete. Thus it may be considered beneficial to use wheels with flat spots on the standard rolling machine. An additional benefit is that the profile of a wheel is easier to control (from a standardization perspective) than a roughness profile.

While use of a flat wheel does have the benefit of generating a louder, more consistent sound, the risk is that the presence of flat spots causes the sound to shift away from sounding like rolling noise and towards sounding like impact noise. This is perhaps analogous to the discussion surrounding the tapping machine on whether or not it is truly representative of footfall noise. Use of flat wheels should remain an option in the case that a sufficiently high signal to noise ratio cannot be achieved with a smooth wheel, or that a roughness profile is considered too difficult to tightly control. However, its drawbacks should still be kept in mind.

The question of what load should be placed on the rolling device may seem to draw parallels to the question of what the hammer momentum (or fall energy) should be for the tapping machine. However, as it has been discovered that the load on the trolley has little influence on the generated sound level, this is actually only a minor design criteria. The sound level has a tendency to decrease slightly with increasing load. The trolley need only be heavy enough to ensure that no rattling or other secondary noise is emitted. Beyond that, it may be kept as light as possible to both increase sound level and reduce the difficulty of moving it around.

\begin{figure}[!htbp]
	\centering
	\subfloat[][One wheel: single point of contact possible]{\includegraphics{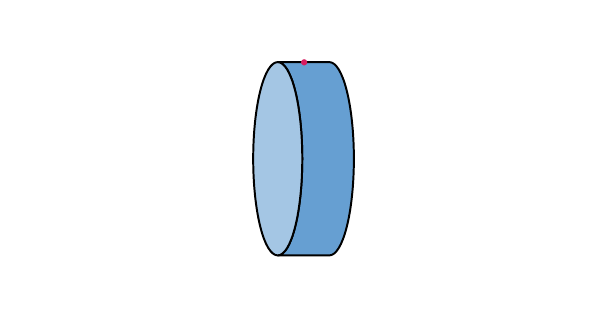}} 
	\subfloat[][Two wheels: single line of contact possible]{\includegraphics{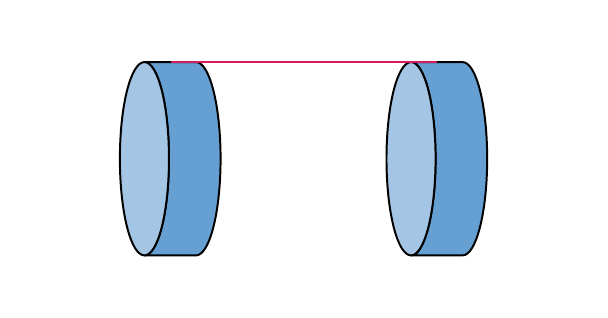}} \\
	\subfloat[][Three wheels: single plane of contact possible]{\includegraphics{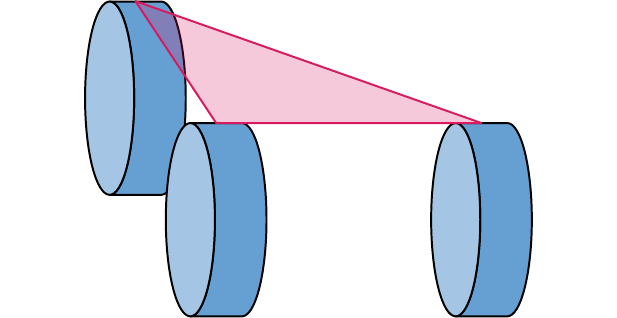}}
	\subfloat[][Four wheels: multiple planes of contact possible]{\includegraphics{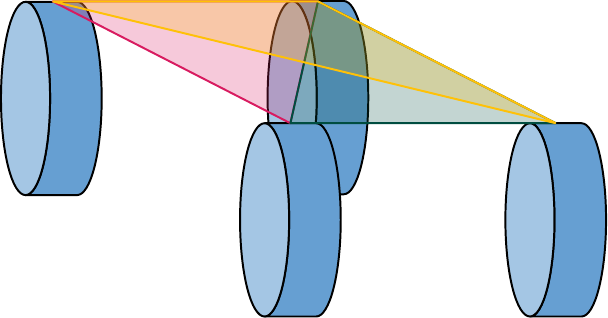}}
	\caption{Different contact scenarios based on the number of wheels. With four wheels and above, it is possible to have more than one plane of contact, introducing wobble.}
	\label{fig:trolley_number_of_wheels}
\end{figure}

In a similar vein, the rolling device should be constructed in such a way so as to ensure all wheels always remain in constant contact with the ground. This can be achieved with an either two or three-wheeled device. As shown in \cref{fig:trolley_number_of_wheels}, moving up to four wheels introduces the possibility that the device may wobble as it rolls, if a perfect plane is not formed between the floor and the four contact points. In the interest of automation, a three wheeled device may be considered preferable over one with two wheels, as the latter would need to be supported by some other means (likely a person) to avoid falling over (technically gyroscopic stabilization could be used on a two-wheel device, but that hardly fulfills the requirement that the design be \emph{simple}). A three-wheeled, motorized design would allow the same flexibility offered by the tapping machine, where it could be set, turned on, and recorded without human interaction. Though considerations would still need to be made to ensure the device does not run into a wall.

Fortunately, it has been shown through rolling noise tests that trajectory and the presence of a human operator provide a negligible change in the radiated sound level. Thus the decision is greatly simplified as to whether to use two or three wheels, and whether to have the trolley automated or manually pushed.

The final variable to consider is speed. The sound level and frequency content of the rolling noise will change with the speed of the rolling device; thus this is a parameter which should be intentionally chosen. Following the goal of replicating a worst case scenario, the speed should be reasonably high such that a high radiated sound level is generated, but not so great as to make the rolling device unstable or difficult to operate reliably. A speed of around 1~m/s, which is slightly below the preferred human walking speed of 1.4~m/s \cite{browningEffectsObesitySex2006}, may be an ideal balance point.

\subsection{Rolling noise measurement methodology} \label{sec:measurement_methodology}
While the focus of this paper is on the development of the device itself, consideration will also need to be made in generating the method used to measure the rolling noise produced by said device. One option would be to use the same ISO standard process already established for measuring impact noise. This has the benefit of being about as straightforward as possible. There is no need to generate a brand new methodology, and rolling noise results may more easily be compared with those of a tapping machine.. However, it is worth pointing out that even the existing method of measuring root mean square (RMS) sound levels in the reception room does not accurately represent the subjective perception of impact noise either \cite{schultzAlternativeTestMethod1976a,hammeSoundTransmissionFloorCeiling1965}. This is not to say that we may take the opinion of, ``well their method has flaws, so it’s okay to put the same flaws in our method too.'' On the contrary, this may serve as an opportunity to correct the mistakes of yesteryear. 

Measuring RMS sound levels should in theory provide a better representation of rolling noise than impact noise, as the former is more homogeneous in nature. Considering the movement of the rolling source across the floor above (contrasted with a stationary tapping machine), directivity may play a role in the perception of the sound in the room below. This is dealt with in the standards surrounding the sound emission of earth moving equipment by having the machine drive through a microphone hemisphere at a constant speed, and measuring the average sound power level of the hemisphere from the time the center of the machine enters one side of the hemisphere to the time it exits the opposite side \cite{ISO639520082008}. Additionally, for testing in reverberant rooms, a microphone on a rotating boom is often used. While the presence of a multi-floor setup with indoor rolling noise prohibits the use of a microphone hemisphere around the rolling source (not to mention the headache that would cause for acousticians, should such a method be adopted), the fact remains that averaging of the recorded sound level over the duration of the rolling event as the device moves from one side of the floor to the other may serve well to remove any discrepancies caused by directivity. Fortunately, this is precisely what is done in ISO 10140-5: either a number of microphones spaced apart or a single microphone on a rotating boom are used in the reception room. Nevertheless, a perceptive test which compares rolling noise recordings captured with different methods could yield beneficial information about how well these methods correspond to what humans actually perceive..

Finally, in order for measurements to remain comparable across laboratory locations, it is advised that a consistent thickness be used for the base concrete slab upon which the floor rolls. As 140~mm is a typical thickness for concrete floors in multi-story buildings, specifying that all rolling noise laboratory measurements be conducted using a base floor of 140~mm concrete would allow for easy comparison across a number of laboratories.

\section{Proposed development process} \label{sec:development_process}
With all the above being said, the proposed development process for a standard rolling machine prototype could be considered as follows:
\begin{enumerate}
	\item Conduct a series of tests on various rolling products in a two-story reverberation chamber in order to gain a wide range of audio samples for the different kinds of rolling noise which may be found in indoor scenarios, thus complimenting the existing indoor rolling test data.
	\item Perform a subjective study to identify how varying the measurement method of the above tests changes the perceived rolling sound compared with what a listener actually hears.
	\item Use the results of the rolling product tests to aid in the development of specific design requirements for the prototype. Choose dimensional and material criteria in such a way as to provide ``worst case scenario'' representation of the range of sound profiles observed.
	\item Develop a theoretical prototype design using a computer-aided design (CAD) software.
	\item Fabricate first prototype. Test the device in a two-story reverberation chamber, comparing the results to those of the previous tested rolling products. Compare both the sound generated by the prototype, as well as the applicability of the measurement method decided upon in step 2.
	\item Revise the prototype design and measurement method as necessary based on the outcomes of the test.
	\item Disseminate results.
\end{enumerate}

\section{Conclusion} \label{sec:AppendixA_conclusion}
This report details a proposal for defining a standard rolling noise machine. Just as the standard tapping machine can be used in any building and on any surface as a way of characterizing and comparing the performance of various floors with respect to impact noise, the development of a standard rolling device would enable the same evaluation and comparison to be made with respect to rolling noise. Throughout the research process, it was discovered that a preference for high repeatability over high accuracy was continuously made in the decisions that went into the standard tapping machine design. These decisions may be used as guidance and inspiration in the future development of a standard rolling machine, with a possible goal being to develop a prototype which may be used as a reference device for conducting rolling noise measurements in buildings.

\section{Acknowledgments} \label{sec:AppendixA_acknowledgements}
The work which went into this report was conducted as part of the Acoutect program in conjunction with NCC during the demonstrator group 3 secondment, November 5--16, 2018. The authors would like to thank NCC for hosting the team during the duration of the secondment: notably Linda Cusumano, Christina Claeson-Jonsson, and Birgitta Berglund. This project has received funding from the European Union’s Horizon 2020 research and innovation programme under grant agreement No 721536.

\bibliographystyle{ieeetr}
\bibliography{RecommendationStandardDevice}

\end{document}